\documentclass[]{spie}  %>>> use for US letter paper
\newif\ifpdf
\ifx\pdfoutput\undefined
\pdffalse % we are not running PDFLaTeX
\else
\pdfoutput=1 % we are running PDFLaTeX
\usepackage[]{epstopdf}
\pdftrue
\fi

\usepackage[]{graphicx}

\title{Applications of quantum message sealing} 

\author{G Gordon Worley III
\skiplinehalf
School of Computer Science, University of Central Florida, Orlando, FL 32816-2362, USA
}

\authorinfo{A version of this paper appeared as SPIE 5815-25, \copyright SPIE 2005.\\Further author information:  E-mail: gworley@cs.ucf.edu, Telephone: 407-823-2341}
\begin{document}
  \maketitle

%%%%%%%%%%%%%%%%%%%%%%%%%%%%%%%%%%%%%%%%%%%%%%%%%%%%%%%%%%%%% 
\begin{abstract}
In 2003, Bechmann-Pasquinucci introduced the concept of quantum seals, a quantum analogue to wax seals used to close letters and envelopes.  Since then, some improvements on the method have been found.  We first review the current quantum sealing techniques, then introduce and discuss potential applications of quantum message sealing, and conclude with some discussion of the limitations of quantum seals.
\end{abstract}

%>>>> Include a list of keywords after the abstract 

\keywords{quantum seals, message sealing, quantum information theory}

%%%%%%%%%%%%%%%%%%%%%%%%%%%%%%%%%%%%%%%%%%%%%%%%%%%%%%%%%%%%%
\section{INTRODUCTION}
\label{sect:intro}

Before the Internet age, most people sent written messages on paper.  These messages could be sealed by melting wax over an overlap of the pages or the cover flap of an envelope.  When dry, the message could only be read by breaking the seal first.  Additionally, the wax was often impressed with an image to indicate authenticity, such as a family crest.  The seal serves two purposes:  it weakly authenticates the sender and provides a way of checking if the message has been opened.  Note that this seal doesn't commit the sender to the content, secure the content, or even guarantee authenticity; it merely provides a simple way of checking if the message has been opened and an indication of who sealed it.

A quantum seal works by hiding some qubits in a quantum message that are not knowable before reading the message.  Then when the message is read, these sealing qubits cause errors in certain bits.  Using error correcting codes, these errors will produce negligible errors in the read message, but will provide a way to check if someone read the message.  As we'll see, quantum sealing depends on the uncertainty principle and the no cloning theorem.\cite{Singh05}

At the time of writing, we know of three published quantum sealing algorithms, each with advantages to the others.  The Bechmann-Pasquinucci scheme appeared first in 2003, shortly followed by Chau in the same year.  In 2005, Singh and Srikanth introduced a modified Bechmann-Pasquinucci scheme that avoids some of the limitations of the original.  Note that although the Bechmann-Pasquinucci scheme was developed for sealing classical bits, Sing and Srikanth show it's possible to seal quantum messages using schemes intended for sealing classical bits,\cite{Singh05} and clearly any classical message can be converted to a quantum message, so in our discussion we'll generally refer to original messages and sealed messages, where a sealed message is always quantum but an original message may be classical or quantum.

\subsection{Bechmann-Pasquinucci∗}

Alice has a message she'd like to seal and send to Bob.  For each bit in this message, she constructs three qubits where two of them are in the reading basis and the other is in a sealing basis, where the reading basis is defined by the communication protocol and the sealing bases are different from the reading basis, mutually unbiased, and kept secret.  Alice then sends the message to Bob or posts it in a public location where they can both access it.\cite{Bechmann-Pasquinucci03}  

\begin{figure}
\begin{center}
\begin{tabular}{c}
\includegraphics{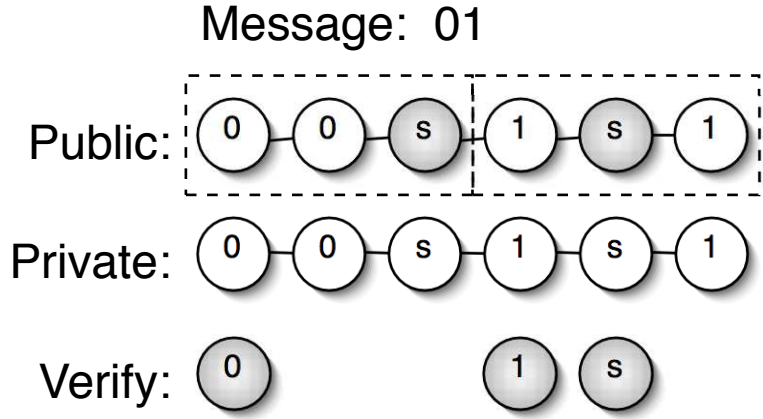}
\end{tabular}
\end{center}
\caption[example] 
%>>>> use \label inside caption to get Fig. number with \ref{}
{ \label{fig:bp}
Consider this possible sealing of the message 01.  First the message is sealed in a $[[3, 1]]_2$ code with one of the qubits a sealing qubit, the details of which are kept secret.  Further, each group of three qubits has a publicly unknown order and value before observation.  In private, though, Alice knows everything about the message.  To verify it, Alice sends Bob some qubits and tells him their location in the message so he can SWAP test them.
}
\end{figure}

To check if the message is sealed, Alice sends Bob a copy of a proper subset of qubits in the sealed message, but he is only told their index, so he knows nothing about their state or if they are message qubits or sealing qubits.  This is possible because Alice created the message and knows the states of all the qubits, so the no cloning theorem is not violated.  Using a SWAP test, Bob can compare the qubits from Alice with the qubits in the sealed message.  If any of these qubits differ, Bob can presume that the message was read; otherwise, the message seal is intact.\cite{Bechmann-Pasquinucci03}  See Fig. \ref{fig:bp} for a graphical example of sealing a message.

This scheme has serious weaknesses and limitations.  As Bechmann-Pasquinucci, D’Ariano, and Macchiavello argue, it's possible to read a Bechmann-Pasquinucci sealed message by reading the qubits collectively rather than individually and leave the seal unbroken.  Also, they prove that a perfect sealing of classical information is impossible if the information is perfectly retrievable.\cite{Bechmann-Pasquinucci05}  Thus although the Bechmann-Pasquinucci quantum sealing scheme is simple to understand and instructive, it has weaknesses that make it impractical for real-world use.

Chau and Singh and Srikanth get around these limitations in different ways.  Chau seals quantum rather than classical information, thus a perfect scheme is not proved impossible, and Singh and Srikanth seal without perfect retrievability, making a perfect sealing possible.

\subsection{Chau}

\begin{figure}
\begin{center}
\begin{tabular}{c}
\includegraphics{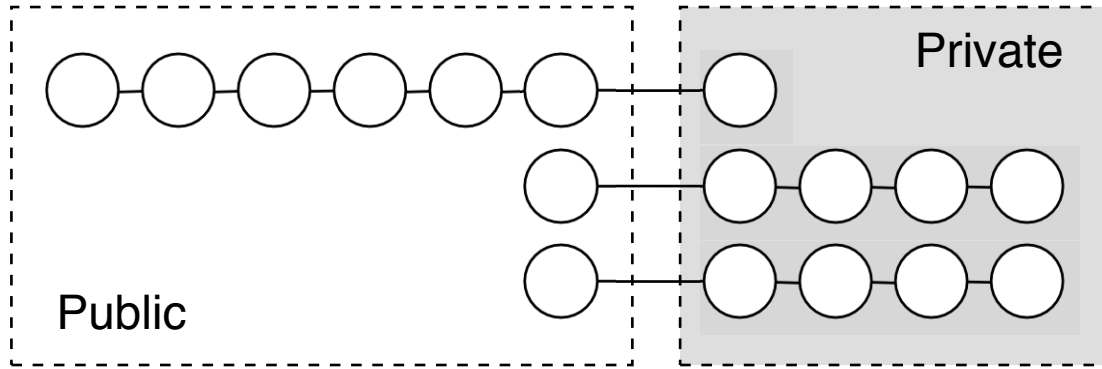}
\end{tabular}
\end{center}
\caption[example] 
%>>>> use \label inside caption to get Fig. number with \ref{}
{ \label{fig:chau}
This is an example taken from Ref. \citenum{Chau03}.  Here a qubit is sealed using a $[[7, 1, 3]]_2$ CSS code and a $[[5, 1, 3]]_2$ stabilizer code.  The qubits in the left box are made public, while the qubits in the right box are kept secret by Alice.
}
\end{figure}

Alice has a message she'd like to send to Bob.  For each qubit $|\phi\rangle$ in this message, she encodes it with a Calderbank-Shor-Steane (CSS) code $[[n, 1, d]]_2, d > 0.11d$ and publicly announces this code.  She then encodes $t \equiv \lfloor\frac{d -1}{2}\rfloor$ copies of $|0\rangle$ through a $[[n', 1, 3]]_2$ stabilizer code and releases $(n - t)$ randomly selected qubits of the encoded message and $t$ qubits of a separate copy of the stabilizer qubits.  Now $n$ bits are public (sent to Bob) and Alice keeps the remaining qubits secret.  This encodes the message such that Bob can read it using the CSS code, but doing so will disturb the unknown qubits in the stabilizer.\cite{Chau03}  See Fig. \ref{fig:chau} for a graphical view of an example sealing.

Alice (or anyone else with the nonpublic qubits) can now verify if the message is sealed by checking for errors in the message sent to Bob.  Since she knows the locations of the stabilizer qubits, she can test if the public qubits are error free.  If they are, the message is sealed.  If not, the message was read.\cite{Chau03}

This scheme is proved unconditionally secure.\cite{Chau03}  However, it has advantages and disadvantages in that Bob cannot tell if the message is sealed.  Although in some cases Alice may wish to keep this information from Bob, other times we want Bob to check if the message is sealed.  Therefore, although Chau is good for sealing some quantum messages, we need another solution to handle others.

\subsection{Singh \& Srikanth}

Singh and Srikanth improve on the Bechmann-Pasquinucci∗ quantum sealing scheme in two ways.  First, they require that message bits consists of the minority of the bits in an error correcting code (relative frequency of message bits less than $\frac{1}{2}$ in the sealed message).  As they argue by the large number theorem, though, the bit will usually be retrievable, and this sidesteps the impossibility of perfect sealing proof.  They also propose sending shares of a secret share bit rather than the bit itself to reduce the chance an unintended reader will gain access to the message bit, thus more strongly authenticating messages than in the other two schemes.  Further, combining quantum encryption and secret sharing, they're able to seal quantum bits.\cite{Singh05}

%%%%%%%%%%%%%%%%%%%%%%%%%%%%%%%%%%%%%%%%%%%%%%%%%%%%%%%%%%%%%
\section{POTENTIAL APPLICTIONS} 
\label{sect:apps}

Quantum seals are similar to, but different from, wax seals on paper.  Here we'll look at some ways to use quantum seals, from direct analogues to surprising results.

\subsection{Message Receipt}

The most common historical use of a wax seal was to close a message so that opening it would announce that someone had read the message.  This acts as a message-read receipt, which is valuable to the sender if the message must be read by a certain time or simply must be read, such as legal documents like court summons and wills.  Using a quantum seal, we can create the digital equivalent of this paper technology.

Consider that Alice wants to send Bob an e-mail and she wants to know if Bob reads this message (assume the e-mail system does not disturb the message until Bob reads it in his mail user agent).  This e-mail may contain important legal information, but it may also be something more personal, like a love note, and Alice doesn't want to mention the note to Bob in person until she is sure that he's read the note.  By putting a quantum seal on the message and, assuming the mail system allows her to verify the seal on a message received by another party, she can check if Bob has read the message.  Of course, she must still be sure to send the e-mail to the right address, because if she sends it to the wrong one she may learn that the message has been read but she won't realize it wasn't read by Bob, much less realize who did read it until she talks to Bob, embarrasses herself, and checks her e-mail logs to discover her error.

\subsection{Protective Packaging}

Sometimes when you buy products they say you may only return them if they are unopened (this is often the case with commercial software to prevent piracy from someone buying the software, making a copy, and then returning it).  These products usually have a simple physical seal on them, but the problem with physical seals is that they can be easily forged and replaced, at least well enough to fool the average retail clerk.  But products could instead contain a quantum sealed message with a mechanism that observes the qubits when the packaging is opened.  When the customer brings the product in for a return, the store could check the quantum seal.  If it's broken, the store may assume the product was opened, thus not acceptable for return.

This may not be a cost effective solution for most products, though, since the cost of the quantum seal would probably exceed the cost of a single act of piracy.  But for expensive software products or other expensive, returnable only if unopened, products, this may provide a better solution than paper, plastic, or wax seals.

Quantum sealing also extends to other packaging situations, like packages sent in the mail.  This would allow the recipient to be sure the package was untampered with and the sender to know if the recipient had opened the package.  If the seal is broken and one of the parties think the delivery protocol was not followed, then they might investigate a misdelivery, such as delivery to the wrong person if the intended recipient reports not opening the package but the sender reports the package opened, or the recipient receives the package but finds the seal already broken, indicating tampering.  Like with sealed products, there are already systems in place to detect such misdeliveries, so the cost of a misdelivery would have to be high enough to warrant the use of a quantum rather than classical seal.

\subsection{Quizzes}

In the academic world, quizzes, tests, and other forms of examination are common tools of assessment.  One of the keys to making these tools fair to all takers is that no one have more advanced knowledge of the material than anyone else.  If the content of a quiz is public knowledge before it is taken, this presents no problems, but if it to remain secret, it would be nice to have a way to be sure it was not accessed.  Quantum seals provide a solution.

Alice writes a quiz for her students and then seals it with a quantum seal.  Bob is one of her students and he would like to know the content of the quiz before he takes it, i.e. he wants to cheat.  Bob hacks Alice's computer and gains access to the quiz but not the seal secret (Alice was careful to store them separately).  Since Bob is trying not to get caught, he can't take Alice's qubits, so he makes a copy of them, but by the no cloning theorem, he must copy the observed qubit states, thus the quiz is distrubed.  The next day, Alice checks the seal and finds it broken, therefore she designs a new quiz, much to the disdain of Bob.  However, Alice has only learned that the quiz was seen.  She doesn't know who saw it or what they did with it.  Note, too, that Alice could have done something more traditional, like encrypt the quiz, but if Bob had stolen it and managed to decrypt it, Alice would not know that the quiz had been seen because she was depending wholly on the encryption to keep it secret.  In the quiz scenario, sealing plus encryption provides a good way to keep the quiz secret until it is given.

This use of seals generalizes to any message we wish to keep secret and wish to detect if the secret may have been learned (if the message is not encrypted, a broken seal will tell us that the secret was learned).  Or it may not be a secret we're trying to protect, but access to the message by intended parties only.

\subsection{Eavesdropping Detection}

Alice wants to send a message to Bob and to know if the message channel's content is being watched by an eavesdropper, Eve (we'll assume Eve can learn without disturbing the message that Alice and Bob are communicating).  Alice could send Bob an encrypted message, but she could not be sure that someone did not intercept it and decrypt it.  Instead, Alice puts a quantum seal on her message and then sends it to Bob.  Upon arrival, they check if the message was read in transit.  If Eve read or tried to copy the message, the no cloning theorem guarantees the seal will have been broken, thus Alice and Bob will know that someone is eavesdropping.  Note that the quantum seal does nothing computational to prevent eavesdropping, but socially prevents it by ensuring that Alice and Bob will know if there is an eavesdropper, although they won't be able to identify who it is, so anyone may eavesdrop anonymously, although awareness of these activities will keep Alice and Bob from revealing information an eavesdropper may want.

Quantum seals could be used to monitor any form of digital communication for eavesdropping, from cell phone conversations to e-mails to network video game signals.  Eavesdropping detection would affect different people in different ways.  For people concerned about their privacy, a quantum seal would allow them to know if they were being observed by a third party and possibly alter their communications to protect secrets (for good or bad).  Police forces may not like this because it will make the job of eavesdropping on criminals more difficult.  Using their power over municipalities, police could defeat quantum seals by disturbing all quantum communications, so Alice and Bob will always believe that someone is eavesdropping whether the police are actually eavesdropping or not.  But disturbing all quantum communications would also prevent most other quantum communication protocols from functioning, which may upset the public enough to keep it from happening.  Thus quantum seals provide a simple, light weight (don't increase the size of messages assuming error correcting codes are already in use) solution to eavesdropping detection.

\subsection{Bit Commitment}

Alice wants to send Bob a committed bit.\cite{Schneier96}  Alice puts a quantum seal on this bit and sends the result to Bob.  Now at any time Alice will be able to check if Bob has read the committed bit and so long as Bob keeps the sealed bit from Alice, she will not be able to change it.  This provides only a loose bit commitment scheme, though, that depends on the social aspects of bit commitment:  Alice doesn't want Bob to see the bit early; Bob doesn't want Alice to change her bit.  For some applications, like gambling results, this may be enough, since a compromised bit can easily be discarded, but for other applications, like trial evidence that is not to be opened until a certain time and should not be tampered with, this is probably not enough since leaked bits cannot be replaced, i.e. you shouldn't commit noncommodity bits with a quantum seal.

All of the applications listed so far have something in common:  all use seals to indicate the binary condition of sealedness.  This suggests we can use quantum seals in any application where a binary condition must be indicated, such as in binary semaphores.

\subsection{Binary Semaphores}

A semaphore is a simple method of solving the critical section problem.  It uses two functions, $P$ and $V$, to try to increment and to decrement a binary semaphore variable $s$, which may be of any data type capable of storing a boolean value.\cite{Wikipedia:Semaphore}  We can use quantum seals to implement binary semaphores by modifying the $P$ and $V$ functions as follows:

\begin{verbatim}
    P(Quantum sealed s):
        wait while seal on s.
        seal s.

    V(Quantum sealed s):
        remove seal from s.
\end{verbatim}

The $P$ function waits so long as $s$ is sealed, corresponding to waiting while another process is in its critical section.  Then when a process finishes its critical section, it calls $V$ to release the semaphore by changing $s$ so that it is no longer sealed (either by observation or by destroying the value).  Back in $P$, $s$ is not sealed, and a process will seal $s$ and enter its critical section, sure it is the only one running in its critical section.

Admittedly, the quantum seal semaphore is a useless construction, both because semaphores are an archaic technique and because integers serve as semaphores without requiring a quantum computer.  However, we include it here to demonstrate the flexibility of quantum seals, in the hopes it may inspire an even less expected application of quantum seals.

\section{LIMITATIONS}

Quantum seals are exciting because they allow us to perform an action on digital information that was not possible before the existence of quantum information theory.  However, in our excitement, we must be careful not to treat quantum seals as a panacea.  Hence, in this last section of our paper, we discuss some of the limitations of quantum seals, potential workarounds, and give some open problems on the limits of quantum seals.

{\bf Quantum seals don't hide information}.  Quantum seals can't, by themselves, perform data hiding procedures, such as steganography and watermarking.  In particular, because the seal only exists in the quantum message and not in the classical message, the hidden information could be easily removed by breaking the seal and passing only the observed classical information.  However, a quantum seal could be applied to a quantum message with hidden data to reveal if a third party tried to retrieve or break the hidden information.

{\bf Quantum seals don't know who broke them.}  When Alice seals a message and later Bob breaks it, the quantum seal, by itself, doesn't reveal the breaker.  In some cases, Alice may have confidence in her knowledge of the identity of the seal breaker, such as when she sends a message to Bob and she knows he receives it, or when she sends a message to Bob and he reports the seal broken before he reads it, viz. the breaker is not Bob.  But in general, a quantum seal reveals no information about the seal breaker, so if Alice wants to know the identity of the breaker, she must use an intrusion detection system.

{\bf Quantum seals are probabilistic}.  Because quantum sealing depends, in part, on quantum uncertainty, there is no guarantee that reading the sealed message will always break the seal, that the message will always be retrievable, or that it will always be possible to accurately assess the state of a quantum seal.  However, as the protocols presented show, it's possible to get tolerable results in all these areas, so that even if there are protocol errors, they will not drastically exceed the system errors.

{\bf Quantum sealed messages must be public.}  When Alice sends Bob a quantum sealed message, if Alice wants to check the seal she must have access to the message she sent to Bob after she's sent it.  Depending on how quantum information networks develop, this may not be practical, limiting Bob and other recipients to checking the seal, thus forcing Alice to rely on the reports of others about the current state of the seal.  While in some uses this may be acceptable, in others Alice needs to be able to reliably check the state of the seal.  Thus this should be a consideration when developing networks to share qubits.

{\bf Current quantum sealing protocols ignore channel noise.}  All of the protocols described use error correcting codes to quantum seal messages.  They all also, however, ignore channel noise, which could disrupt the quantum seal or even the message.  So an open problem is to develop a quantum sealing protocol that survives noisy channels.

{\bf Quantum seals don't authenticate messages.}  Quantum seals have limited ability to authenticate messages.  Although all schemes involve a certain level of authentication because checking if the message is sealed requires qubits from the sender, none of them provide good protocols for sender authentication.  So another open problem is to develop a good combination of quantum seal and quantum signature to produce something superior to even a direct analogue of classical sealing.\cite{Gottesman01}

{\bf Quantum seals don't hide messages from public view}  Despite the `sealing' terminology, quantum seals don't keep messages from the public, except perhaps by the social implications of breaking a quantum seal.  But, by combining quantum seals with quantum cryptography, we seal a message and hide it from public view.  There are at least two ways to accomplish this:  encrypt a message and then seal it or seal a message and then encrypt it.  In Ref. \citenum{Singh05}, Singh and Srikanth show a way of encrypting a message and then sealing it.  This allows you to know if someone tried to decrypt a message, but not if they decrypted it.  What would be even better is a protocol for encrypting a quantum sealed message, such that decrypting the message and then reading it would break the seal, revealing that the message had been decrypted.

As we've shown, quantum message sealing cannot do everything we might like it to, but it has the power to do things we could never do before.

%%%%%%%%%%%%%%%%%%%%%%%%%%%%%%%%%%%%%%%%%%%%%%%%%%%%%%%%%%%%%
\acknowledgments     %>>>> equivalent to \section*{ACKNOWLEDGMENTS}       
 
Thanks to Dan C. Marinescu for introducing me to quantum information theory and assisting me with my early research.  Thanks to Charles E. Hughes for supporting this work by keeping my teaching load light.  Additional thanks to Rene Peralta for showing me bit commitment while I was writing this paper.

%%%%%%%%%%%%%%%%%%%%%%%%%%%%%%%%%%%%%%%%%%%%%%%%%%%%%%%%%%%%%
%%%%% References %%%%%

\bibliographystyle{spiebib}   %>>>> makes bibtex use spiebib.bst
\bibliography{appquantseal}   %>>>> bibliography data in report.bib

\begin{thebibliography}{1}

\bibitem{Singh05}
S.~K. Singh and R.~Srikanth, ``Quantum seals,'' January~2005.
\newblock To appear in Physica Scripta, arXiv:quant-ph/0410017.

\bibitem{Bechmann-Pasquinucci03}
H.~Bechmann-Pasquinucci, ``Quantum seals,'' {\em International Journal of
  Quantum Information}~{\bf 1}(2), pp.~217--224, 2003.
\newblock arXiv:quant-ph/0303173.

\bibitem{Bechmann-Pasquinucci05}
H.~Bechmann-Pasquinucci, G.~M. D’Ariano, and C.~Macchiavello, ``Impossibility
  of perfect quantum sealing of classical information,'' January~2005.
\newblock To appear in International Journal of Quantum Information,
  arXiv:quant-ph/0501073.

\bibitem{Chau03}
H.~F. Chau, ``Sealing quantum message by quantum code,'' August~2003.
\newblock arXiv:quant-ph/0308146.

\bibitem{Schneier96}
B.~Schneier, {\em Applied Cryptography}, John Wiley \& Sons, Inc., New York,
  1996.

\bibitem{Wikipedia:Semaphore}
``Semaphore (programming),'' 2005.
\newblock Accessed on 28 February 2005 at
  http://en.wikipedia.org/wiki/Semaphore\_(programming).

\bibitem{Gottesman01}
D.~Gottesman and I.~L. Chuang, ``Quantum digital signatures,'' November~2001.
\newblock arXiv:quant-ph/0105032.

\end{thebibliography}

\end{document}